\documentclass{aa}
\usepackage{graphicx,natbib,amsmath,amssymb}
\usepackage{txfonts}
\bibpunct[, ]{(}{)}{;}{a}{}{,}

\begin{document}

   \title{The supernova 2003lw associated with X-ray flash 031203\thanks{The observations from the Danish 1.5 m Telescope were supported by the Danish
Natural Science Research Council through its Center for Ground Based
Observational Astronomy (IJAF).}\fnmsep\thanks{Based, in part, on observations
made with the Nordic Optical Telescope, operated on the island of La Palma
jointly by Denmark, Finland, Iceland, Norway, and Sweden, in the Spanish
Observatorio del Roque de los Muchachos of the Instituto de Astrof\'{\i}sica de
Canarias.}
}

   \author{B.~Thomsen\inst{1}
           \and
           J.~Hjorth\inst{2}
           \and
           D.~Watson\inst{2}
           \and
           J.~Gorosabel\inst{3}
           \and
           J.~P.~U.~Fynbo\inst{1,2}
	   \and
	   B.~L.~Jensen\inst{2}
	   \and
	   M.~I.~Andersen\inst{4}
	   \and
	   T.~H.~Dall\inst{5}
	   \and
	   J.~R.~Rasmussen\inst{1}
	   \and
	   H.~Bruntt\inst{1}
	   \and
	   E.~Laurikainen\inst{6}
	   \and
	   T.~Augusteijn\inst{7}
	   \and
	   T.~Pursimo\inst{7}
	   \and
	   L.~Germany\inst{5}
           \and
           P.~Jakobsson\inst{2}
           \and
           K.~Pedersen\inst{2}
          }

   \offprints{B.~Thomsen, email: \texttt{bt@phys.au.dk}}

   \institute{Department of Physics and Astronomy, University of Aarhus, Ny Munkegade, DK-8000 \AA rhus C
              \and
              Niels Bohr Institute, Astronomical Observatory, University of Copenhagen, Juliane Maries Vej 30, DK-2100 Copenhagen \O, Denmark
              \and
	      Instituto de Astrof\'{\i}sica de Andaluc\'{\i}a (IAA-CSIC), Apartado de Correos, 3004, E-18080 Granada, Spain
                            \and
	      Astrophysikalisches Institut Potsdam, D-14482 Potsdam, Germany
              \and
	      European Southern Observatory, Alonso de C\'ordova 3107, Casilla 19001, Santiago 19, Chile
              \and
	      Department of Physical Sciences, University of Oulu, Box 3000, FIN-90014 Oulu, Finland
              \and
	      Nordic Optical Telescope, Apartado 474, E-38700 St.~Cruz de La Palma, Canary Islands, Spain
              	                   }

   \date{Received / Accepted }

   \abstract{The X-Ray Flash (XRF), 031203 with a host galaxy at
             $z=0.1055$, is, apart from GRB\,980425, the closest
             $\gamma$-Ray Burst (GRB) or XRF known to date.  We have
             monitored its host galaxy from 1--100\,days after the burst. 
             In spite of the high extinction to the source and the bright
             host, a significant increase and subsequent decrease has been
             detected in the apparent brightness of the host, peaking
             between 10 and 33\,days after the GRB.  The only convincing
             explanation is a supernova (SN) associated with the XRF,
             SN2003lw.  This is the earliest time at which a SN signal is
             clearly discernible in a GRB/XRF (apart from SN1998bw). 
             SN2003lw is extremely luminous with a broad peak and can be
             approximately represented by the lightcurve of SN1998bw
             brightened by $\sim0.55$\,mag, implying a hypernova, as
             observed in most GRB-SNe.  The XRF--SN association firmly links
             XRFs with the deaths of massive stars and further strengthens
             their connection with GRBs. The fact that SNe are also
             associated with XRFs implies that \emph{Swift} may detect a
             significant population of intermediate redshift SNe very soon
             after the SN explosions, a sample ideally suited for detailed
             studies of early SN physics.
     \keywords{gamma rays: bursts -- supernovae: general
               }
   }

   \maketitle

%
%
\section{Introduction\label{introduction}}
It is now firmly established that at least some long-duration $\gamma$-ray
bursts (GRBs) are accompanied by the contemporaneous explosion of a
supernova \citep[SN,
e.g.][]{2003Natur.423..847H,2003ApJ...591L..17S,2003A&A...406L..33D,1998Natur.395..670G},
consistent with expectations for some models of GRBs involving the collapse
of massive stars \citep{2001ApJ...550..410M,1999ApJ...524..262M}.  The lack
of large numbers of SN--GRB associations may be explained, at least
in part, by the difficulty of obtaining the optical spectra of SNe with
redshifts usually greater than unity, against the combined backgrounds of
the fading afterglow and the host galaxy.

X-Ray Flashes (XRFs), a class of very soft bursts, was discovered with
\emph{BeppoSAX} \citep{2001grba.conf...16H}. They are intense, short-lived
flashes of soft X-rays of extragalactic origin
\citep{2003ApJ...599..957B,2003astro.ph.11050S,2004astro.ph..2085P}, and may
be defined by a larger X-ray than $\gamma$-ray fluence in the burst
\citep[$S_{\rm X}/S_\gamma>1$,][]{2003astro.ph.12634L}.  The similarity in
the durations of XRFs and GRBs
\citep{2001grba.conf...16H,2003A&A...400.1021B}, the continuum of spectral
properties observed between the two classes \citep{2003astro.ph.12634L},
their cosmological origins in each case
\citep{2003ApJ...599..957B,2003astro.ph.11050S}, and the similarity of their
optical and X-ray afterglows
\citep{2004astro.ph..2240F,2004astro.ph..1225W}, makes it seem probable that
XRFs and GRBs have a similar origin. While GRBs and XRFs are located at
cosmological distances, few have been located at redshifts $<0.3$.  They
are: GRB\,030329 at $z=0.1685$
\citep[associated with SN2003dh,][]{2003Natur.423..847H,2003ApJ...591L..17S},
XRF\,020903 with its probable host galaxy at $z=0.251$
\citep{2003astro.ph.11050S}, GRB\,980425 probably associated with
SN1998bw at $z=0.0085$ \citep{1998Natur.395..670G} and recently, the XRF
referred to as GRB\,031203 at $z=0.1055$
\citep{2004astro.ph..1225W,2004astro.ph..2085P}.  Of these, XRF\,020903 had
a very low peak spectral energy and a low luminosity
\citep{2004ApJ...602..875S,2003astro.ph.11050S} and GRB\,980425 had an
extraordinarily low luminosity \citep{1998Natur.395..663K}.  That
GRB\,031203 was in fact an XRF was discovered because of the detection of a
transient, outwardly moving ring of X-ray emission surrounding the afterglow
\citep{2004ApJ...603L...5V}.  This was interpreted as reflection of the original burst
event and early afterglow off dust sheets in the Galaxy, from which strong
lower limits on the prompt soft X-ray fluence were obtained \citep{2004astro.ph..1225W}.

The high extinction toward GRB\,031203 \citep[$(\,\rm
E(B-V)\approx1$,][]{1998ApJ...500..525S,2004astro.ph..2085P}, though
instrumental in allowing the detection of the dust reflection halo, also
hampered attempts to follow the afterglow at optical wavelengths and no
optical or infrared afterglow was detected. The location of the burst is
therefore determined from X-ray and radio detections.  Both of these locate
GRB\,031203 unambiguously on a sub-luminous, blue, and strongly star-forming
galaxy at fairly low redshift \citep[$z=0.1055$,][]{2004astro.ph..2085P},
where the probability of a chance association with such a galaxy is not very
significant \citep{2004astro.ph..1225W,2004astro.ph..2085P}.

Strong evidence for the association of XRFs with the deaths of massive stars
was present in the lightcurve of XRF~030723 \citep{2004astro.ph..2240F}, but
the analysis of that burst was complicated by the lack of a redshift.
Because of the low redshift there was considerable interest in attempting to
discover a SN associated with GRB\,031203, in particular since it has been
found to be an XRF and the host galaxy was therefore monitored independently
by a number of groups in order to find the photometric variability that
would indicate a SN \citep[e.g. this
paper;][]{2003GCN..2544....1B,2003GCN..2545....1T,2004astro.ph..3510C,2004astro.ph..3608G}. This
would be the first SN associated with an XRF with a spectroscopic redshift,
firmly establishing the association of XRFs with the deaths of massive stars
and confirming the suspected link between XRFs and GRBs.

In Sect.~\ref{observations} we describe I-band imaging observations of the
host galaxy of GRB\,031203 (HG\,031203) over the first 100\,days since the
burst, and in Sects.~\ref{results} and \ref{discussion} the discovery of a SN
\citep[named SN2003lw, ][]{2004IAUC.8308....1T} associated with the XRF and the
implications of this discovery. This paper supersedes an earlier preliminary
report on some of these observations \citep{2003GCN..2493....1H}.

A cosmology where $H_0=75$\,km\,s$^{-1}$\,Mpc$^{-1}$, $\Omega_\Lambda = 0.7$ and
$\Omega_{\rm m}=0.3$ is assumed throughout.

%
%
\section{Observations and data reduction\label{observations}}
HG\,031203 was observed in the I-band with the DFOSC instrument in imaging
mode (0.396 arcsec/pixel) on the Danish 1.5\,m telescope, La Silla, and with
StanCam (0.176 arcsec/pixel) on the 2.56\,m Nordic Optical Telescope, La
Palma.

The main obstacle to obtaining accurate photometry was strong reflection of
the light from a very bright star off optical surfaces in the camera lens
\citep[this was the cause of the early report of a flat lightcurve
in][]{2003GCN..2493....1H}. The telescope pointing was changed between
exposures in the usual way in order to eliminate the influence of pixel
defects by taking the median of the sky-aligned images. This, however, is
not the best way to proceed when dealing with large reflections. Instead, we
decided to do straight aperture photometry on the individual images that
were free of reflections near the host galaxy. Flat field exposures were
obtained of the twilight sky for each night 
and bias subtraction and flat-fielding were done in the standard way.

\begin{figure}
  \includegraphics[width=\columnwidth,clip=]{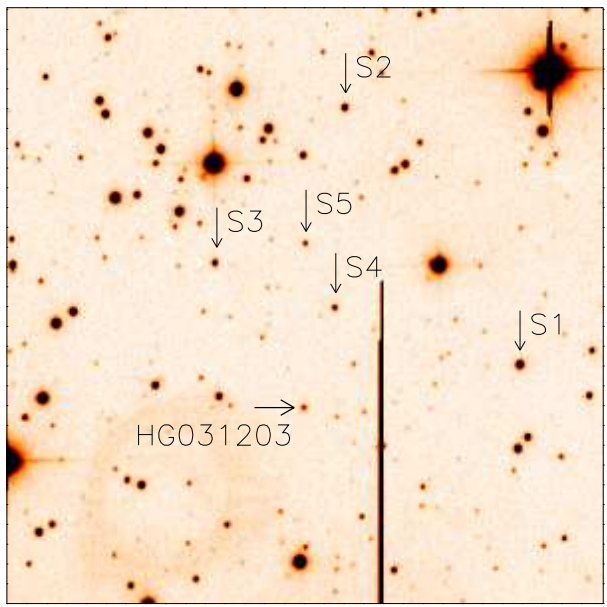}
  \caption{I-band image of the field of HG\,031203.  North is up, east to
           the left; the image is 2\arcmin\ on a side.  The positions of the
           host galaxy and of the comparison stars are noted.  Blooming from
           a bright star, (outside this image) is visible as a N-S streak
           and a ring caused by reflection is evident just to the southeast
           of HG\,031203.
         }
  \label{fig:field}
\end{figure}

An isolated comparison star (S1), 2.5\,mag brighter than HG\,031203, and
four fainter stars (S2--S5) in the vicinity of the host, were chosen to
evaluate the precision of our aperture photometry (Fig.~\ref{fig:field}).
Photometry was carried out using the \texttt{DAOPHOT} package supplemented
by \texttt{daomatch} and \texttt{daomaster} kindly supplied by Peter Stetson
(priv.\ comm.), and used an aperture of 1.98\arcsec\ radius, while the sky
level was estimated using an annular aperture with inner and outer radii of
3.96\arcsec\ and 5.94\arcsec\ respectively. The full width at half maximum
(FWHM) of the seeing was 0.9--1.1\arcsec.

Weighted average instrumental magnitudes were calculated for each night
using weights based on the standard errors supplied by the aperture
photometry routine in the \texttt{DAOPHOT} package; these errors are based
on photon and read noise only. 
The instrumental magnitudes obtained with StanCam were colour corrected to
the DFOSC instrumental magnitudes by demanding identical relative photometry
derived from nearly simultaneous exposures obtained by StanCam and DFOSC. 
This colour-correction was 0.13\,mag in the case of HG\,031203.
 
The photometry of stars S2--S5 and the host galaxy, relative to the
comparison star S1, is given in Table~\ref{tab:photometry}. The standard
error for the magnitude difference is also given, assuming photon and read
noise only. As the pointing was changed between exposures, we can expect
some flat-fielding errors independent of magnitude; the principle
uncertainty is, however, related to the influence of reflections from the
bright star -- such additional background errors are expected to be larger
for fainter stars. In addition to the bright ring-shaped reflections the sky
background appears somewhat non-uniform (Fig.~\ref{fig:field}), but it is
difficult to judge if this non-uniformity is due to scattered light or
flat-fielding errors.  The variability of the four stars and the host galaxy is tested
by calculating the $\chi^2$ per degree of freedom (DOF, $\chi^2_\nu =
\chi^2$/DOF) for a constant flux fit, which is also given in
Table~\ref{tab:photometry}. The fact that for the comparison stars,
$\chi^2_\nu\geq1.5$, suggests that other sources of error must remain in
addition to the photon noise.  These are probably caused by a combination of
small scale background variations and unavoidable flat-fielding errors.
Judging from the standard errors and the $\chi^2_\nu$ values given in
Table~\ref{tab:photometry}, the combined errors could hardly exceed
0.02\,mag. The very high value of $\chi^2_\nu$ for the host galaxy strongly
suggests the existence of an intrinsically variable source superposed on the
galaxy.

The zeropoint used for the absolute photometry was tied to that of
\citet{2004astro.ph..3510C}.

%
%

\section{Results\label{results}}

\begin{figure}
  \includegraphics[bb=70 375 530 694,width=\columnwidth,clip=]{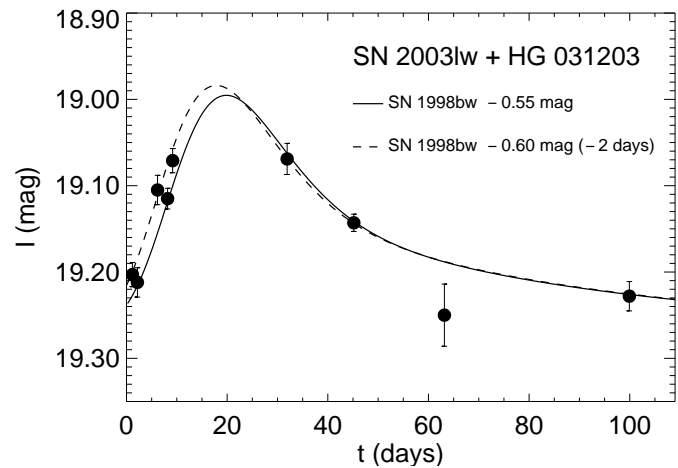}
  \caption{The I-band lightcurve of HG\,031203 (uncorrected for extinction).
           The characteristic supernova ``bump'' is apparent.  The reddened
           (A$_{\rm I} = 2.14$\,mag) and scaled ($-0.55$\,mag) intrinsic
           lightcurve of SN1998bw, with the flux of the galaxy ($I_{\rm
           host} = 19.27$) added, is plotted for comparison, starting at the
           time of the XRF (solid line) and starting two days prior to the
           XRF (scaled by $-0.6$\,mag, dashed line).  The brightened
           SN1998bw lightcurve is quite an accurate template for the data.
         }
  \label{fig:supernova}
\end{figure}

The lightcurve of HG\,031203 is clearly variable and rises and then falls in
a manner characteristic of a SN superimposed on a host galaxy
(Fig.~\ref{fig:supernova}).  While it is possible that the first two
data\-points may be related to the afterglow of the XRF, it is clear from the
shape and timing of the bump that there was a SN (2003lw) in HG\,031203
contemporaneous with the XRF. This is the first SN associated with an XRF
\citep[with known redshift, see][]{2004astro.ph..2240F}. On
Fig.~\ref{fig:supernova} we plot SN1998bw as it would appear in HG\,031203
(assuming $I_{\rm host} = 19.27$\,mag) and it gives a reasonable
approximation to the lightcurve if we allow it to be brighter by 0.55\,mag. 
However, if SN2003lw and GRB\,031203 were simultaneous, the rise appears
marginally faster than SN1998bw; alternatively, placing the start of
SN1998bw template up to two days prior to the XRF is consistent with the
observations (Fig.~\ref{fig:supernova}).
It is already apparent from the lightcurves of SNe associated with GRBs/XRFs
(011121, 021211, 030329, 030723), that SN1998bw is not a universal template,
with some being faster \citep[2003dh,][]{2003Natur.423..847H} or having an
early peak in the near-infrared, and fading more quickly e.g.\ 2001ke
\citep{2003ApJ...582..924G} or XRF\,030723 \citep{2004astro.ph..2240F}. 
Given the variation in GRB-associated SN lightcurves, it is slightly
surprising that SN2003lw follows the brightened SN1998bw template fairly
well.

Whether SN2003lw is intrinsically brighter than SN1998bw, depends entirely
on the extinction to the SNe.  Using the values measured from the Balmer
decrement, \citet{2004astro.ph..2085P} found the total E(B$-$V)=$1.17\pm0.1$,
(somewhat higher than, but consistent with, the values obtained from
Galactic dust-maps \citep[E(B$-$V)=1.04,][]{1998ApJ...500..525S}, with
R$_V=3.1$ \citep{1989ApJ...345..245C} giving a good fit to the ratio of Balmer line fluxes
\citep{2004astro.ph..2085P}.  Therefore, the total I-band extinction toward
HG031203 is $A_{\rm I} = 2.14\pm0.2$\,mag and is the value used here. This
means that SN2003lw is likely to have been $\sim0.55$\,mag brighter at peak
than SN1998bw. Since we use a very high value for the total extinction
toward SN1998bw, $A_{\rm I} = 0.12$,\footnote{This is in fact an upper limit
to the extinction in SN1998bw, based on the non-detection \ion{Na}{i} D
lines in high-resolution spectra \citep{2001ApJ...555..900P}.} this estimate
may be slightly low. It seems unlikely that the total extinction to SN2003lw
is much less than $A_{\rm I} = 1.4$\,mag, a value $>3\,\sigma$ lower than
that measured using the Balmer line ratios mentioned above and consistent
with the lowest estimate of Galactic extinction in this direction
\citep[and references therein]{2004astro.ph..2085P}.  Using these limits,
SN2003lw must therefore have been at least as bright as SN1998bw. This
implies a high mass of $^{56}$Ni produced in the explosion
\citep{1998Natur.395..672I}, and together with the similarity with the
lightcurve of SN1998bw, suggests that SN2003lw was a hypernova.

The existence of SN2003lw in HG\,031203, suspected soon after the GRB
\citep{2003GCN..2486....1B} and suggested again much later
\citep{2003GCN..2544....1B}, also now appears to have been confirmed
spectroscopically \citep{2003GCN..2545....1T}.  Our results appear
consistent with those preliminary reports.  Recently
\citet{2004astro.ph..3510C} with more extensive temporal coverage but larger
photometric uncertainties than reported here, suggest that the fast rise
(Fig.~\ref{fig:supernova}) noted above and a broader peak implies a later
maximum than we have inferred.

\begin{table*}
 \caption{Comparative photometry relative to the star S1. Relative
          photometric uncertainties in the last significant digit are listed
          in parentheses after their respective values.  The last column
          gives the reduced $\chi^2$ of the fit to a constant flux value. 
          Seeing corresponds to the full width at half-maximum of point
          sources in the image.  The data were obtained with DFOSC on the
          Danish 1.5\,m telescope except that marked with a $\dagger$, where
          StanCam on the NOT was used.}
 \label{tab:photometry}
 \begin{tabular}{lcccccccccr}
  \hline\hline
  & \multicolumn{9}{c}{Days since XRF (3.92 Dec.\ 2003, UT)}	& \\
 & 1.36      & 2.29      & 6.43       & 8.28       & 9.33       & 32.19$^\dagger$     & 45.34      & 63.45     & 100.12     & \\[3pt]
\hline
Exp.\ (s): & $4\times600$ & $3\times600$ & $3\times600$ & $5\times420$ & $3\times600$ & $6\times 300$  & $5\times600$ & $4\times600$ & $5\times600$& \\
Seeing:      & 1.09\arcsec & 1.03\arcsec &  0.95\arcsec &  0.87\arcsec &  0.85\arcsec &  0.83\arcsec &  0.98\arcsec &  1.05\arcsec & 1.00\arcsec & \\[3pt]
  \hline
     Object & \multicolumn{9}{c}{Relative Photometry} & $\chi^2_\nu$\\
\hline
       S2 &  1.000(5)  &  0.997(5)  &  1.013(6)  &  1.009(4)  &  1.004(5)  &  0.994(7)  &  1.004(3)  &  1.01(1)  &  0.992(5)  & 1.71 \\
       S3 &  1.396(6)  &  1.386(7)  &  1.388(8)  &  1.378(6)  &  1.378(6)  &  1.390(9)  &  1.407(4)  &  1.42(1)  &  1.391(6)  & 3.29 \\
       S4 &  1.923(9)  &  1.96(1)  &  1.92(1)  &  1.921(8)  &  1.93(1)  &  1.94(1)  &  1.929(6)  &  1.91(2)  &  1.90(1)  & 3.15 \\
       S5 &  2.36(1)  &  2.37(2)  &  2.41(2)  &  2.39(1)  &  2.39(1)  &  2.35(2)  &  2.371(9)  &  2.37(3)  &  2.40(2)  & 1.50 \\
HG\,031203 &  2.47(1)  & 2.48(2) &  2.37(2)  &  2.38(1)  &  2.34(1)  &  2.33(2)  &  2.41(1)  &  2.52(4)  &  2.49(2)  & 15.26\\
\hline
 \end{tabular}
\end{table*}

%
%
\section{Discussion\label{discussion}}

The SN associated with GRB\,031203 is a confirmation of an expectation that
XRFs and GRBs are essentially two ends of the continuum of cosmic
high-energy bursts
\citep[e.g.][]{2002A&A...390...81A,2004ApJ...602..875S,2004astro.ph..1225W}
that result from the destruction of massive stars. It has been posited that
the lower apparent luminosities and lower peak energies of XRFs may both be
related to the fact that these bursts are viewed at larger off-axis angles
than GRBs \citep{2002ApJ...570L..61G,2002ApJ...571L..31Y,2004astro.ph..1142Y,2003astro.ph..8389Z}.
XRFs are typically found at lower redshifts than GRBs, probably because of
their lower overall luminosities \citep{2002A&A...390...81A,%
2003A&A...407L...1A,2002ApJ...571L..31Y}. Since there is
no obvious reason to expect that the SNe associated with XRFs are less
luminous optically than the SNe associated with GRBs, it seems likely that
proportionally more XRF-SNe than GRB-SNe will be found if the limiting factor
to discovering GRB/XRF-SNe is simply the very large distances involved.

Now that the association with SNe is secure, the question arises of what the
general characteristics of GRB-SNe are.  The fact that the absolute
magnitudes of the SNe of GRB\,980425, 011121, 021211, 030329 and now 031203
are all within a magnitude of each other (the SN associated with 021211
being the faintest and 031203 the brightest), centred near $M_{\rm B}\sim-19.5$\,mag \citep{1998Natur.395..670G,2002ApJ...572L..45B,2003ApJ...582..924G,2003A&A...406L..33D,2003Natur.423..847H,2003astro.ph.12594L}
when the mean $M_{\rm B}$ of Type Ib/c SNe \citep[the SN type of
SN1998bw/GRB\,980425 and
2003dh/GRB\,030329,][]{1998IAUC.6918....1P,2003ApJ...599L..95M,2003ApJ...599..394M}
is $-17.12$\,mag, with a standard deviation in the magnitudes of 0.74
\citep{2002AJ....123..745R,1990AJ....100..530M} is intriguing.  The fact
that there is some evidence for a bimodal distribution in the absolute
magnitudes of Type I\,b/c SNe
\citep{2002AJ....123..745R,1990AJ....100..530M} with mean $M_{\rm B}=-19.77$\,mag and
standard deviation of 0.33 for the brighter group merely heightens the
interest. For the moment the statistics are not good enough to go beyond the
statement that it is possible that only the brightest SNe can be observed
above the afterglow and galaxy light, and therefore we sample the high end
of the population preferentially.  However, it does not seem unreasonable to
suggest that such a bimodal distribution may indeed exist in SNe\,Ib/c and
that this more luminous population subset is associated with GRBs, or
with the specific population of stars that produce GRBs.

GRB\,031203 was a relatively faint \emph{Integral} burst with a peak flux of
only 1.3$\times$10$^{-7}$ erg cm$^{-2}$ s$^{-1}$
\citep{2003GCN..2460....1M}. In terms of peak $\gamma$-ray luminosity,
GRB\,031203 is one of the faintest localised so far. The \emph{Swift}
satellite, with its high $\gamma$-ray sensitivity and large field of view,
should detect many more faint GRBs than previous missions. While some of
these will be at high redshift, there should be a significant population of
intrinsically faint XRFs (and GRBs) at modest redshift, similar to
GRB\,031203, that will also be detected. These will provide excellent
targets to study the earliest phases in the evolution of Type Ib/c SNe from
minutes to months after the burst, especially considering that most of
these, contrary to GRB 031203, will be relatively unextincted (had
GRB\,031203 not been close to the plane of the Galaxy, the SN peak magnitude
would have been $m_{\rm V}\sim19$\,mag). In this way \emph{Swift} may
unintendedly open an entirely new research field within SN physics, allowing
extremely early access to, and a substantial increase in the rate of
detections of, Type Ib/c SNe.

%
%
\section{Conclusions\label{conclusions}}
We have monitored the host galaxy of the XRF, GRB\,031203 in the
near-infrared, from 1--100\,days following the burst.  In spite of the
bright host galaxy and high extinction, we have discovered positive evidence
of a SN, peaking $\sim20$\,days after the XRF and can clearly trace the
early SN rise.  At $z=0.1055$, this is the closest GRB/XRF-associated SN
discovered so far, after SN1998bw, and the first SN associated with an XRF
with known redshift. This confirms the strong case for an association
between XRFs and SNe found in XRF\,030723 \citep{2004astro.ph..2240F}. The
SN appears to have a somewhat higher peak luminosity than observed in
SN1998bw, but the lightcurve is otherwise fairly similar, implying that the
SN accompanying GRB\,031203 was a hypernova.  It is likely that
\emph{Swift} will detect a significant population of faint bursts like
GRB\,031203 and hence allow the study of (Type Ib/c) core-collapse SN at
much earlier times than what has been possible so far; this may have a
substantial impact on SN research.

\begin{acknowledgements}
We acknowledge benefits from collaboration within the EU FP5 Research
Training Network, `Gamma-Ray Bursts: An Enigma and a Tool'. This work was
also supported by the Danish Natural Science Research Council (SNF).
\end{acknowledgements}

\small
\bibliography{mnemonic,grbs}

\end{document}